\documentclass[usenatbib]{aat} 

\usepackage{gensymb}
\usepackage{flushend}
\begin{document}
\setcounter{page}{23}
\nametom{3(1), \pageref{firstpage}--\pageref{lastpage} (2022)}
\title{Misalignments in NGC 1068}
\author{Pierre Vermot\inst{1,2}}
\institute{Astronomical Institute, Academy of Sciences, Bo\v{c}n\'{\i} II 1401, Prague, Czech Republic\\ \email{pierre.vermot@asu.cas.cz} \and LESIA, Observatoire de Paris, Université PSL, CNRS, Sorbonne Université, Université de Paris, 5 place Jules Janssen, 92195 \\ Meudon, France  } 
\date{Submitted on December 1, 2021}
\titlerunning{Misalignments in NGC 1068}
\authorrunning{Pierre Vermot} 

\maketitle
\label{firstpage}

\begin{abstract} 
NGC 1068 is a nearby Active Galactic Nucleus (AGN) of type 2, meaning that its accretion disk is hidden behind a large amount of foreground extinction. Observations at several wavelengths have revealed various disk-like structures around the nucleus, all possibly part of the putative torus responsible for the obscuration of the AGN. The paper presents results based on GRAVITY/VLTI interferometric observations in the near-infrared, which provide very high angular resolution, and gives insights into the geometry of the innermost region of the torus. The 3D orientation of the structure is surprising in several aspects, as it is misaligned with other disks present around the nucleus, and leaves a clear line of sight toward the central source.
\end{abstract}

\section{Introduction}

AGN are observed with various characteristics in terms of luminosity and spectral properties. A large amount of these differences can be explained by the Unified Model of AGN originally proposed by \citet{Antonucci1993}, which states that a structure surrounding the central source causes drastic extinction for some positions of the observer. This parsec-scale obscuring structure (commonly called the \emph{torus}) is assumed to be located in the equatorial plane of the jet, and to be the continuation of the large-scale accretion flow.

The AGN of NGC 1068 is among the easiest to observe, as it is relatively close to us ($D_L = 14.4$ Mpc, with 1$''$ corresponding to 70 pc), and is located in the center of a face-on spiral galaxy. As a consequence, the object has been observed by many instruments, and a wealth of information is available at various wavelengths and spatial scales. It is known to be a type 2 AGN, meaning that its central source is obscured by a large amount of foreground material. The following list presents some of the main structures present in the nucleus of NGC 1068; from large to small spatial scales:
\begin{itemize}
    \item The Narrow Line Region, which is an ionized outflow observed and characterized through spatially resolved emission line Doppler shifts in the UV and IR \citep{Das2006, Poncelet2008}, extends up to several hundred parsecs from the nucleus. It appears to have a hollow bicone shape with a revolution axis almost in the plane of the sky ($i \sim$ 5--10$\degree$) and oriented with PA $\sim 30\degree$.
    \item An extended scattering structure, observed through polarimetric imaging with SPHERE and described by \citet{Gratadour2015}. With dimensions of $20 \times 60$~pc, it appears to be seen edge-on, and is oriented with \mbox{PA $\sim120\degree$}.
    \item A molecular disk, observed through CO and HCN emission lines with ALMA \citep{Garcia2016, Gallimore2016, Impellizzeri2019}. With a typical spatial scale of $\sim 10$ pc, it is aligned along PA $\sim110\degree$ and seems to have an inclination comprised between $34$ and $66\degree$. Interestingly, this disk is counter-rotating with respect to both the external regions of the galaxy and the inner maser disk (described below).
    \item An edge-on maser disk observed with the VLBA, with an inner radius $r_{\mathrm{in}}\sim0.65$ pc and an outer radius $r_{\mathrm{out}}\sim1.1$~pc, oriented along PA $\sim135\degree$ \citep{Greenhill1996}.
\end{itemize}

The different structures described above are good examples of the general cylindrical geometry assumed for AGN, with an outflow along the polar axis and cold, obscuring material in the equatorial plane. However, as one can notice in the $i$ and PA values given above, these structures are only approximately aligned with these two preferred directions.

The next section presents the geometrical properties of the hot dust located within the inner radius of the maser disk, based on the analysis of very high angular resolution observations obtained with GRAVITY, the near-infrared instrument of the VLTI. The orientation of this hot dust with respect to other components of the nucleus will be discussed in the following sections.

\section{Observation and results}

This study is based on a GRAVITY observation of NGC~1068, performed during the night of November 20, 2018 \citep[][ESO programmes IDS 0102.B-0667, 0102.C-0205 and 0102.C-0211]{GRAVITY2017}. GRAVITY is an instrument of the VLT interferometer providing high angular resolution information in the K band.

Despite being one of the most luminous AGN in the near-infrared, NGC 1068 is in the low-end of GRAVITY observable sources and the instrument had to be used in a peculiar way, with all the flux from the source being injected into the Fringe Tracker, and the science channel being left empty.

The observation provided several interferometric observables: the squared visibility, the closure phase; and the total and coherent spectra. With the Fringe Tracker, these measurements are available with a low spectral resolution $R\sim22$, corresponding to four spectral channels in the K band. The maximum spatial frequency probed by the interferometer is $\sim 60\ M\lambda$, corresponding to an angular resolution slightly above $3$ mas.

To interpret the interferometric measurements, the radiative transfer simulation code MontAGN \citep{Grosset2018} was used to produce flux calibrated images of physically realistic heated dusty disks in the spectral channels of the instrument. The Fourier Transforms of these images were then used to compute synthetic interferometric observables: squared visibility (with wavelength dependence), closure phase (idem), coherent spectrum in arbitrary units and total calibrated spectrum (sum of the resolved and unresolved components), which are then directly compared to the observation.

\begin{figure}
\centering
\includegraphics[width=95mm]{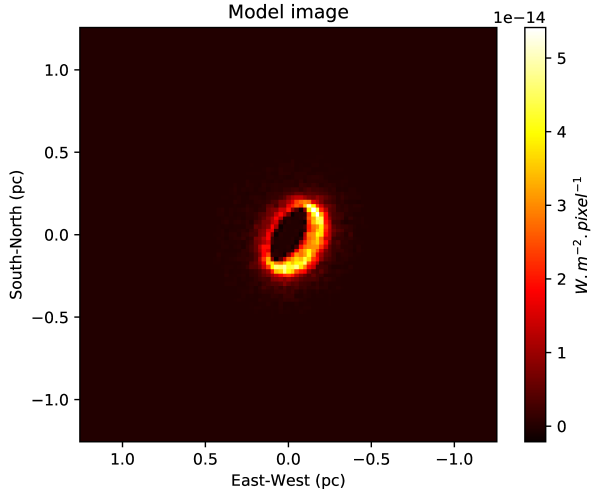} 
\caption{K band image of the best model for the inner hot dust of NGC 1068. 1 parsec corresponds to $\sim15$ mas, and the resolution of the interferometer is $\sim3$ mas.}
\label{fig:image}
\end{figure}

\begin{figure*}
\centering
\includegraphics[width=0.95\textwidth]{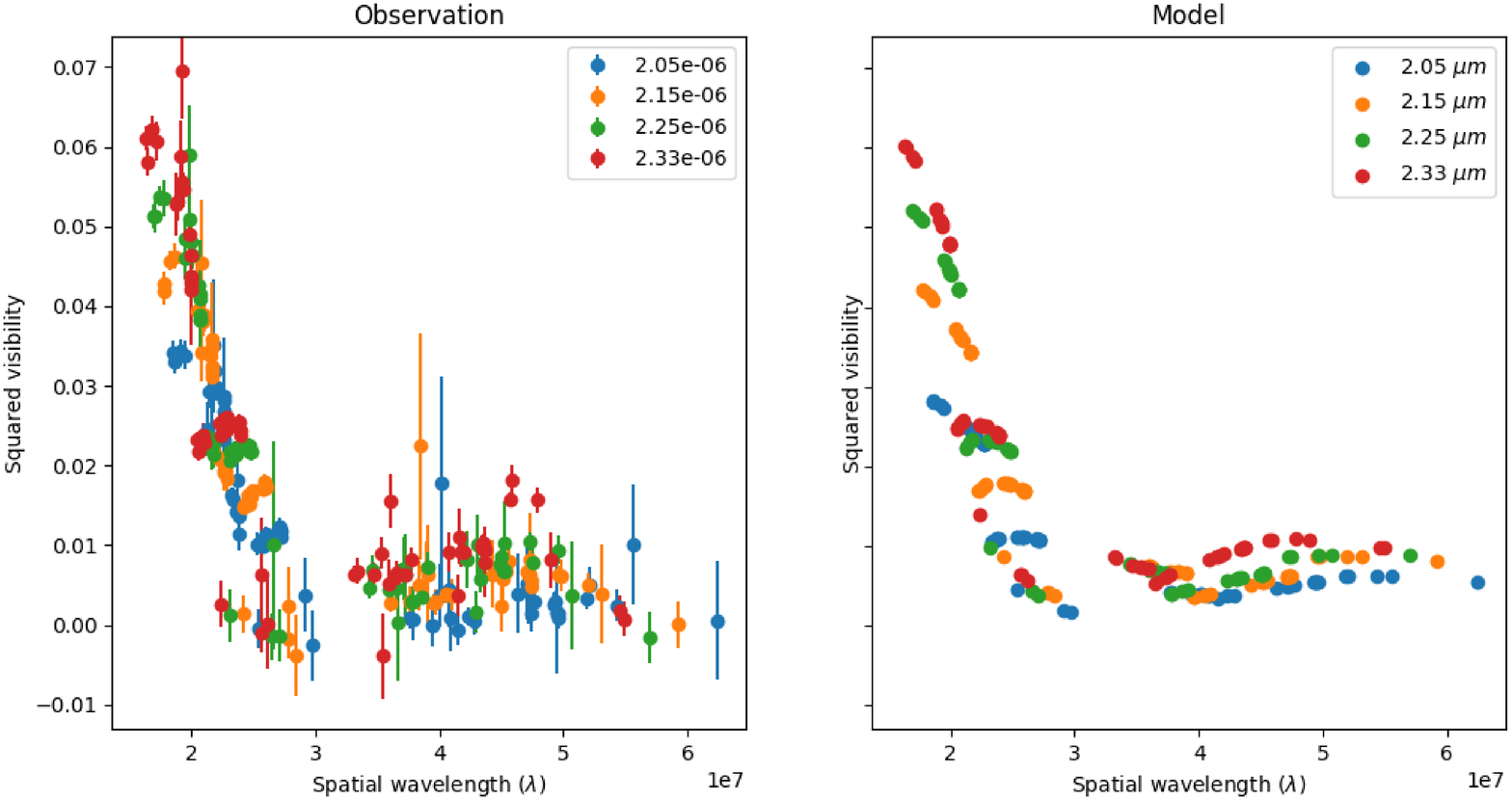} 
\includegraphics[width=0.95\textwidth]{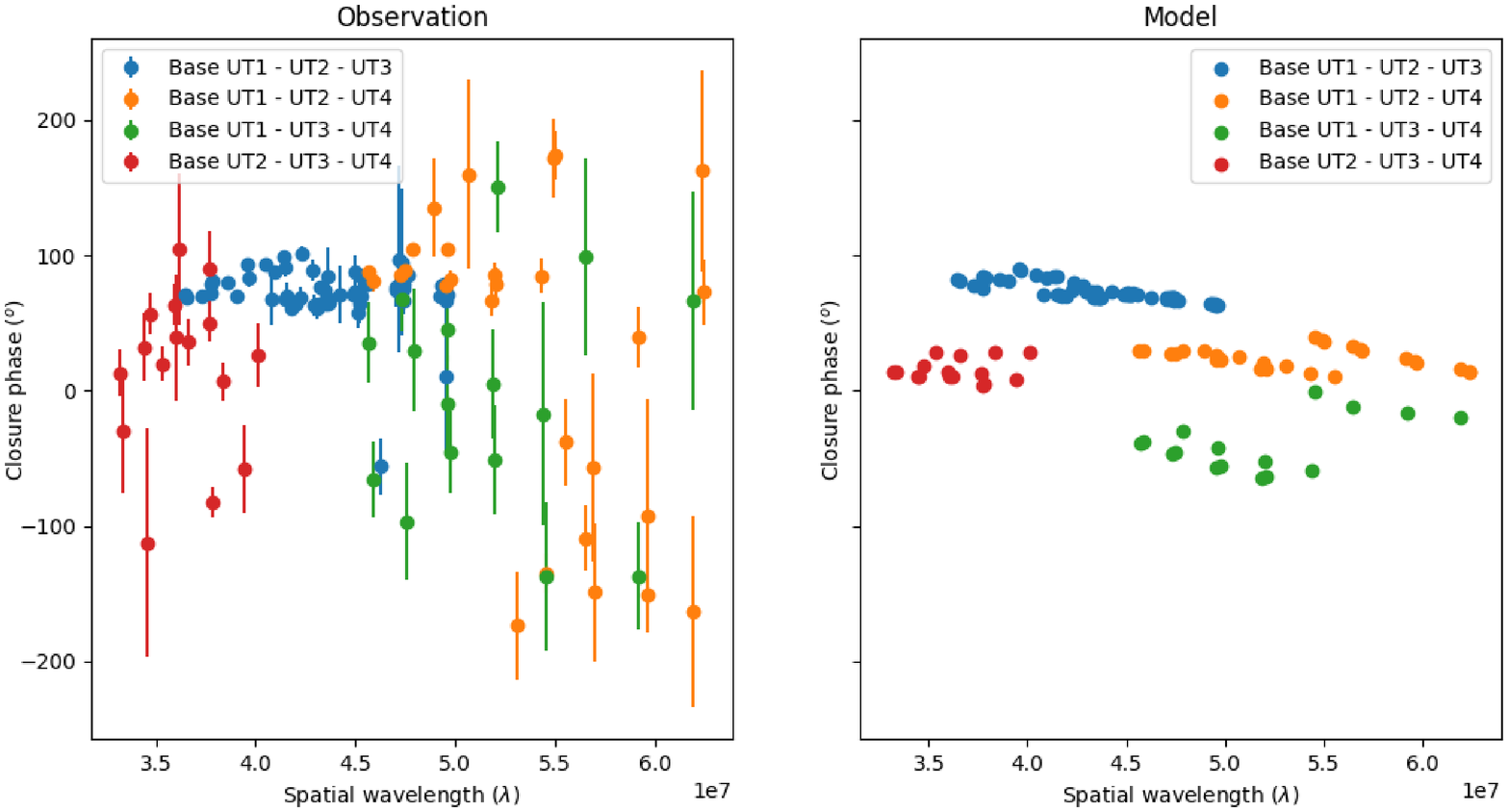} 
\caption{Comparison of the model with observed quantities. Top panel: observed and modeled squared visibility. Bottom panel: observed and modeled closure phase.}
\label{fig:comparison}
\end{figure*}

The hot dust structure was modeled by a thick uniform disk of graphite grains, whose inner edge is heated up to its sublimation temperature by a central UV-X source. Several parameters are fully explored -- the inner radius, the opening angle, the inclination, the position angle, and the density -- and an optimal solution is found, which is in fair agreement with all the interferometric observables. An image of the best model is provided in Fig. \ref{fig:image}, a summary of its parameters in Table \ref{tab:best_model}, and a comparison with the observed visibility and closure phase in Fig. \ref{fig:comparison}. A more detailed description of the modeling and comparisons with the observation is provided by \citet{Vermot2021}.

\begin{table}
	\centering
	\caption{Best model parameters.}
	\label{tab:best_model}
	\begin{tabular}{lccr} 
      Parameter & Value \\
      \hline
      Inner radius & $0.21_{-0.03}^{+0.02}$ pc \\
      Half-opening angle & $21_{-8}^{+8}$ deg \\
      Inclination & $44_{-6}^{+10}$ deg \\
      Density & $10_{-5}^{+10}$ cm$^{-3}$ \\
      PA &  $150_{-13}^{+8}$\,deg \\
	\end{tabular}
\end{table}

\section{Non-obscuring hot dust}

The first remarkable feature of our best model is its inclination, which leaves a clear line of sight for the observer toward the central source. This result is surprising since it is known for a long time that NGC 1068 is a type 2 object, meaning that its accretion disk is obscured by a large amount of foreground extinction. In the commonly accepted description of AGN, it is expected to have this obscuring material distributed in an equatorial structure, orthogonal to the outflow. The hot dust ring does not follow this trend.

The inner radius of the modeled hot dusty disk is constrained from the geometric information obtained with the GRAVITY observations. Assuming that the inner edge of the hot dusty disk is at the sublimation temperature, its radius constrains the UV-X luminosity of the central source, which will be absorbed by the dust and re-emitted in the infrared. Hence, the measure of the inner radius and opening angle of the dusty disk should provide a realistic estimate of the IR magnitude of the source. However, this is not the case, and a high foreground extinction $A_V \geq 50$ needs to be included in the model to account for the observed K band magnitude. This result is supported by the fact that the luminosity of the central source in the model (between $3.9\times 10^{38}$ W and $6.4\times10^{38}$ W), constrained mostly by the size of the inner radius, matches well the previous estimate from independent measurements, ranging from a few $10^{37}$ W to a few $10^{38}$ W \citep{Pier1994, BlandHawthorn1997, Kishimoto1999, Gallimore2001}.

These two conclusions indicate that the inner edge of the dusty torus is not responsible for the extinction toward the central accretion disk, but is rather itself surrounded by a large amount of obscuring material.

\section{Misalignments}

The second remarkable feature of this hot dusty ring model is its misalignment with respect to the previously mentioned structures. Compared to the extended scattering torus, it is not seen edge-on, and the estimations of their PA differ by $30\degree$. The inclination of the molecular torus could be compatible, but their PA differ even more. At last, our model is similar to the maser disk in terms of PA but greatly differs in terms of inclination, with the maser detection requiring a disk closely aligned to the line of sight.

Since these structures are also misaligned between each other, and that several methods used to model the GRAVITY observations gave similar geometrical parameters (geometric modeling in \citet{Vermot2021} and image reconstruction in \citet{GRAVITY2020}), one can reasonably assume that this new misalignment is not an artifact resulting from the interpretation of the interferometric data. 

Geometrically, these observations at different wavelengths could probe different radii of a warped large scale-disk. However, this wouldn't explain the counter-rotation observed between the molecular and maser disks, and no sign of warping is present in the maser observations. Hence, it can be concluded that these different observations are not probing a unique structure, but several ring-like objects intricated around the nucleus. Such structures could result from the tidal disruption of molecular clouds surrounding the nucleus, as seen in the Milky Way. 

\acknowledgements The author is grateful to the referee for a constructive report which helped to significantly improve the paper, as well as the Astronomical Institute of the
Czech Academy of Sciences for support by the Czech Science Foundation Grant 19-15480S and by the project RVO:679858.

\bibliographystyle{aat}
\bibliography{biblio}

\label{lastpage}
\end{document}